\documentclass[preprintnumbers,aps,prb,preprint]{revtex4}

\usepackage{graphicx}
\usepackage{amsmath}
\usepackage{amssymb}

\begin{document}

\title{Stability criterions of an oscillating tip-cantilever system in dynamic
force microscopy}

\author{L.Nony$^{\ast, 1}$, R.Boisgard$^2$, J.-P.Aim\'{e}$^2$\\
 \small{$^1$ L2MP, UMR CNRS 6137,
 Universit\'{e} d'Aix-Marseille III \\
 Facult\'{e} des Sciences de Saint-J\'{e}r\^{o}me, 13397 Marseille Cedex 20, FRANCE}\\
 \small{$^2$ CPMOH, UMR CNRS 5798, Universit\'{e} Bordeaux I\\
 351, cours de la Lib\'{e}ration, 33405 Talence Cedex, FRANCE}\\
 \small{$^\ast$ To whom correspondence should be addressed; E-mail:
 laurent.nony@l2mp.fr}
 Submitted April 12, 2001, \textbf{published in European Physical Journal B  24, pp221-229 (2001)}}

\begin{abstract}
This work is a theoretical investigation of the stability of the non-linear behavior of an oscillating tip-cantilever system used in dynamic force
microscopy. Stability criterions are derived that may help to a better understanding of the instabilities that may appear in the dynamic modes,
Tapping and NC-AFM, when the tip is close to a surface. A variational principle allows to get the temporal dependance of the equations of motion
of the oscillator as a function of the non-linear coupling term. These equations are the basis for the analysis of the stability. One find that
the branch associated to frequencies larger than the resonance is always stable whereas the branch associated to frequencies smaller than the
resonance exhibits two stable domains and one unstable. This feature allows to re-interpret the instabilities appearing in Tapping mode and may
help to understand the reason why the NC-AFM mode is stable.
\end{abstract}

\maketitle

\section{Introduction}

During the last decade, Dynamic Force Microscopy (DFM) used in the Tapping
mode has been found as a suitable tool to investigate surface morphology,
particularly of soft materials and has been widely used to investigate a
very large range of samples including polymers \cite{Stocker96,Magonov97},
biological materials \cite{Rivetti96,Shlyakhtenko99} or organic layers \cite
{Schwartz92,Barrat92,Vallant98}. In recent years, the DFM was developed
around the Non-Contact Atomic Force Microscopy (NC-AFM) mode and has shown
that contrasts at the atomic scale could be achieved on semiconductors and
insulators surfaces \cite
{Geissibl95,Sugarawa95,Kitamura96,Bammerlin96,Schwarz99,99p003}.
Experimental\ and theoretical features dedicated to the description of these
two dynamic modes have been widely discussed in previous papers \cite
{Albrecht91,Anczycowsky96,Wang97,Boisgard98,NCAFMOsaka98,Sasaki99,NCAFMPONTRESINA99}%
. It was shown that the large sensitivity of the oscillating tip-cantilever
system (OTCS) was based on the large value of its quality factor and on its
non-linear dynamics in the vicinity of the surface \cite{Aime2_99,Aime3_99}.
Even if the precise origin of the NC-AFM images contrast remains an open
question, Tapping and NC-AFM results are the consequence of the same
non-linear behavior of the OTCS so that their differences are purely
technical. The aim of this article is to show from a theoretical point of
view that the non-linear dynamics of the OTCS leads to various stability
domains that may help to a better understanding of the way the instabilities
appear in Tapping and the reason why the NC-AFM mode, while being so
sensitive, keeps, in most of cases, a stable behavior.

\bigskip

The experimental study of the stability of the non-linear behavior of the
OTCS can be performed through the study of approach-retract curves, from
which are deduced the variations of the oscillations properties as a
function of the tip-sample distance \cite{Boisgard98,Nony99}. But the
ultimate goal of the approach-retract curves is to extract the surface
properties from amplitude and phase variations (Tapping) or from frequency
shift and damping signal (NC-AFM). In other words, the measure of the
evolution of the amplitude with a fixed drive frequency and a drive
amplitude or the one of the frequency shift are two complementary ways to
investigate the non-linear behavior of the OTCS. Thus in this work, the
analysis of the stability of the oscillations will be done from the study of
the distortion of the resonance peak of the oscillator due to the coupling
with the surface.

\bigskip

The paper is organized as follow. The first part is dedicated to a
description of the non-linear behavior of the OTCS at the proximity of the
surface. To do so, a specific theoretical frame giving the explicit temporal
dependance of equations of motion of the OTCS is developed. The description
is then based on the equations of the stationary state of the OTCS that can
be interpreted either for the Tapping or for the NC-AFM. In the second part,
these equations are used to analyze the stability of the stationary state.
The third part is a discussion of the results obtained and the way they can
be interpreted for DFM experiments.\newpage

\section{Non-linear behavior of the OTCS at the proximity of a surface}

\subsection{General features}

\begin{figure}[h]
\includegraphics[width=10cm]{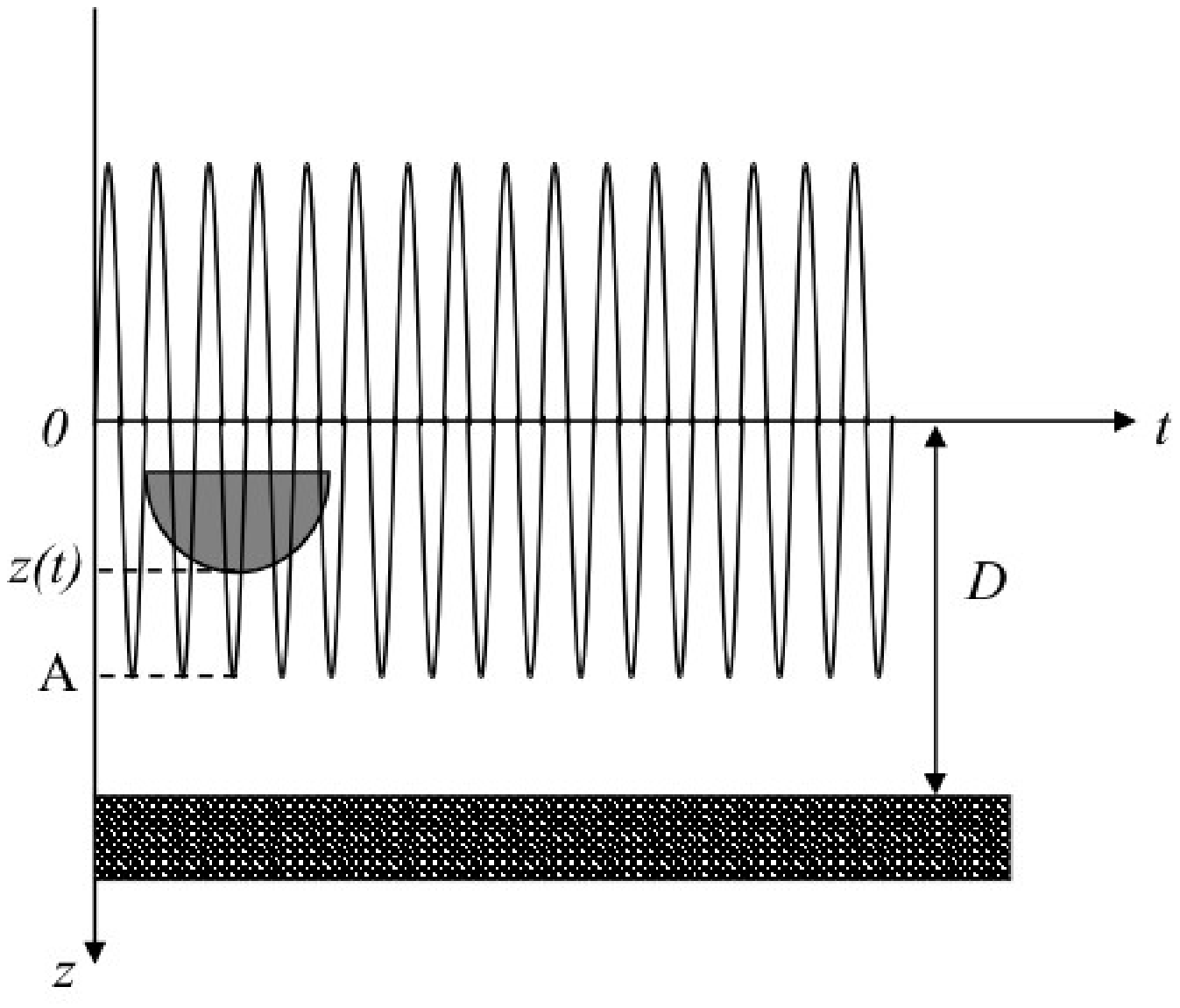}\\
  \caption{Sketch of the OTCS and notations used.}\label{figschemageom}
\end{figure}

The sketch of the OTCS and notations used in the present work are shown in
fig.\ref{figschemageom}. When a tip oscillates above a surface, several
situations can occur. If the oscillation amplitude $A$ is smaller than the
distance $D$ between the surface and the equilibrium position of the OTCS,
the tip never touches the surface. On the contrary, for oscillation
amplitudes larger than $D$, the tip exhibits intermittent contact
situations. The time during which the tip touches the surface depends on
several factors, among which, the local stiffness of the surface is a key
parameter. For example, for very soft materials, the indentation depth can
be as much as the oscillation amplitude \cite{Marsaudon00}, thus the tip
spends half of the oscillation period into the substrate. For hard
materials, the time during which the tip touches the surface is negligible
compared to the oscillation period. Thus, when intermittent contact
situations occur, various assumptions are required to describe the contact
between the tip and the surface \cite{Nony99,Garcia99,Dubourg00}. For the
present purpose, such an analysis introduces useless complications,
therefore we only focus on the non-contact situations. When uniquely
non-contact situations take place, the tip never touches the surface during
the whole oscillation period so that the interaction between the tip and the
surface can be described with a simple expression.

The attractive coupling force between the tip and the sample is assumed to
derive from a sphere-plane interaction involving the disperse part of the
Van der Waals potential \cite{Israelachvili92}. The differential equation
describing the position of the tip's apex as a function of time, $z\left(
t\right) $, is given by~:

\begin{equation}
m^{\ast }\ddot{z}\left( t\right) +\frac{m^{\ast }\omega _{0}}{Q}\dot{z}%
\left( t\right) +k_{c}z\left( t\right) =\mathcal{F}_{exc}\cos \left( \omega
t\right) -\triangledown V_{int}\left[ z\left( t\right) \right] \text{,}
\end{equation}
with~:

\begin{equation}
V_{int}\left[ z\left( t\right) \right] =-\frac{HR}{6\left[ D-z\left(
t\right) \right] }  \label{equpotVdWdisp}
\end{equation}
$\omega _{0}$, $Q$, $m^{\ast }$ and $k_{c}=m^{\ast }\omega _{0}^{2}$ are
respectively the resonance pulsation, quality factor, effective mass and
cantilever' stiffness of the OTCS. $\mathcal{F}_{exc}$ and $\omega $ are the
external drive force and drive pulsation. $H$, $R$ and $D$ are the Hamaker
constant, the tip's apex radius and the distance between the surface and the
equilibrium position of the OTCS (see fig.\ref{figschemageom}).

\bigskip

Different ways can be used to observe the non-linear behavior of the
oscillating nanosphere at the vicinity of a surface. The most informative is
to record several resonance curves as a function of the distance $D$ \cite
{Gleyzes91,Wang97,Sasaki99,Aime3_99}. Another possibility is to keep the
oscillation at a given frequency with a fixed excitation amplitude and to
record variations of the oscillation amplitude and of the phase as a
function of $D$. This latter measurement is readily done by performing
approach curves with the Tapping mode \cite
{Anczycowsky96,Bachelot97,Wang97,Nony99}. As noted in the introduction, the
difference between Tapping and NC-AFM is purely technical but, from a
practical point of view, a description of the variation of the oscillating
behavior showing the correspondence between the two modes is not
straightforward. For example, the NC-AFM mode is probably the simplest way
to describe a DFM experiment \cite{Albrecht91}, but the non-linear behavior
appears in a quite subtle way while it is immediately observed with the
Tapping mode.

\bigskip

The present section is divided into three parts. The first one details the
specific theoretical frame for the obtention of the equations of motion in
amplitude and phase of the OTCS. A coarse-grained method gives the equation
describing the time evolution of the stationary state of the OTCS as a
function of the coupling term between the tip and the surface. This allows
to compute the stability of the stationary state. The next part is a
description of the distortion of the resonance peak as a function of the
distance. This provides the basis of the discussion about the oscillating
behavior and the stability of the branches which is detailed in section \ref
{sectionstability}. This also provides an easier way to discuss experimental
measurements as those obtained in Tapping and NC-AFM modes. As an example,
we use the evolution of the resonance peak to discuss typical variations of
the amplitude and phase in Tapping mode when the drive frequency is slightly
below the resonance one \cite{Nony99}. These variations, regularly observed,
are the most obvious experimental evidences of the non-linear dynamics of
the OTCS. The third part deals with the analysis of the resonance frequency
shift observed in NC-AFM and the way its stability can be interpreted.

\subsection{Theoretical frame}

We search a solution to the temporal evolution of the OTCS by using a
variational solution based on the principle of least action. Even though
this approach exploits the same physical concepts than the one which had led
to the coupled equations in amplitude and phase of the stationary state of
the OTCS \cite{Boisgard98, Aime3_99,Nony99}, it appears to be more general
since here, the temporal dependance is explicitly obtained. We start from
the definition of the action of the OTCS coupled to an interaction
potential~:
\begin{equation}
S=\int_{t_{a}}^{t_{b}}\mathcal{L}\left( z,\dot{z},t\right) dt\text{,}
\end{equation}
where $\mathcal{L}$ is the Lagrangian of the system \cite{Nony99}~:

\begin{eqnarray}
\mathcal{L}\left( z,\dot{z},t\right) &=&\mathcal{T}-\mathcal{V}+\mathcal{W}
\label{equLagrangien} \\
&=&\frac{1}{2}m^{\ast }\dot{z}\left( t\right) ^{2}-\left[ \frac{1}{2}%
k_{c}z\left( t\right) ^{2}-z\left( t\right) \mathcal{F}_{exc}\cos \left(
\omega t\right) +V_{int}\left[ z\left( t\right) \right] \right] -\frac{%
m^{\ast }\omega _{0}}{Q}z(t)\underline{\dot{z}}\left( t\right)  \nonumber
\end{eqnarray}
Due to the large quality factor, we assume that a typical temporal solution
is on the form~:
\begin{equation}
z\left( t\right) =A\left( t\right) \cos \left[ \omega t+\varphi \left(
t\right) \right] \text{,}  \label{equfonctionharmonique}
\end{equation}
where $A\left( t\right) $ and $\varphi \left( t\right) $ are assumed to be
slowly varying functions with time compared to the period $T=2\pi /\omega $.
The underlined variables of $\underline{\dot{z}}\left( t\right) $~:
\[
\underline{\dot{z}}\left( t\right) =\underline{\dot{A}}\left( t\right) \cos %
\left[ \omega t+\underline{\varphi }\left( t\right) \right] -\underline{A}%
\left( t\right) \left[ \omega +\underline{\dot{\varphi}}\left( t\right) %
\right] \sin \left[ \omega t+\underline{\varphi }\left( t\right) \right]
\text{,}
\]
are calculated along the physical path, thus they are not varied into the
calculations \cite{Goldstein80}.\ While among the trial functions solutions
that can be chosen the present one is not the more general, the principle of
least action ensures that functionals of this type are the best.

The equations of motion in amplitude and phase of the OTCS are obtained by
considering the following coarse-grained operation. Let's assume a long
duration $\Delta t=t_{b}-t_{a}$ with $\Delta t\gg T$ and calculate the
action as a sum of small pieces of duration $T$~:
\begin{equation}
S=\sum_{n}\int_{nT}^{\left( n+1\right) T}\mathcal{L}\left( z,\dot{z}%
,t\right) dt=\sum_{n}\left( \frac{1}{T}\int_{nT}^{\left( n+1\right) T}%
\mathcal{L}\left( z,\dot{z},t\right) dt\right) T=\sum_{n}\mathcal{L}_{e}T
\end{equation}
$\mathcal{L}_{e}$ is the mean Lagrangian during one period and\ appears as
an effective Lagrangian for a large time scale compared to the period. Owing
to the quasi-stationary behavior of the amplitude and the phase over the
period, the effective Lagrangian is calculated by keeping them constant
during the integration. The calculations give~:
\begin{eqnarray}
\mathcal{L}_{e}\left( A,\dot{A},\varphi ,\dot{\varphi}\right) &=&\frac{%
m^{\ast }}{4}\left[ \dot{A}^{2}+A^{2}\left( \omega +\dot{\varphi}^{2}\right) %
\right] -\frac{k_{c}A^{2}}{4}+\frac{\mathcal{F}_{exc}A\cos \left( \varphi
\right) }{2}-\frac{1}{T}\int_{0}^{T}V_{int}\left[ z\left( t\right) \right] dt
\nonumber \\
&&-\frac{m^{\ast }\omega _{0}}{2Q}\left[ A\underline{\dot{A}}\cos \left(
\varphi -\underline{\varphi }\right) -A\underline{A}\left( \omega +%
\underline{\dot{\varphi}}\right) \sin \left( \underline{\varphi }-\varphi
\right) \right]
\end{eqnarray}
Note that the effective Lagrangian is now a function of the new generalized
coordinates $A$, $\varphi $ and their associated generalized velocities $%
\dot{A}$, $\dot{\varphi}$. At this point, remembering that the period is
small regardless to $\Delta t=t_{b}-t_{a}$ during which the total action is
evaluated, the continuous expression of the action is~:
\begin{equation}
S=\int_{t_{a}}^{t_{b}}\mathcal{L}_{e}\left( A,\dot{A},\varphi ,\dot{\varphi}%
\right) d\tau \text{,}
\end{equation}
where the measure $d\tau $ is such that $T\ll d\tau \ll \Delta t$.

Applying the principle of least action $\delta S=0$ to the functional $%
\mathcal{L}_{e}$, we obtain the Euler-Lagrange equations for the effective
Lagrangian~:
\begin{equation}
\frac{d}{d\tau }\left( \frac{\partial \mathcal{L}_{e}}{\partial \dot{A}}%
\right) -\frac{\partial \mathcal{L}_{e}}{\partial A}=0\hspace{0.5cm}\text{and%
}\hspace{0.5cm}\frac{d}{d\tau }\left( \frac{\partial \mathcal{L}_{e}}{%
\partial \dot{\varphi}}\right) -\frac{\partial \mathcal{L}_{e}}{\partial
\varphi }=0
\end{equation}
The amplitude and phase equations of motion of the OTCS coupled to an
interaction potential $V_{int}\left[ z\left( t\right) \right] $ are~:

\begin{equation}
\left\{
\begin{array}{c}
\ddot{A}=\left[ \left( \dfrac{\omega }{\omega _{0}}+\dot{\varphi}\right)
^{2}-1\right] A-\dfrac{1}{Q}\dot{A}+\dfrac{\mathcal{F}_{exc}\cos \left(
\varphi \right) }{k_{c}}-\dfrac{\omega }{\pi k_{c}}\dfrac{\partial I\left(
A,\varphi \right) }{\partial A} \\
\ddot{\varphi}=-\left( \dfrac{2\dot{A}}{A}+\dfrac{1}{Q}\right) \left( \dfrac{%
\omega }{\omega _{0}}+\dot{\varphi}\right) -\dfrac{\mathcal{F}_{exc}\sin
\left( \varphi \right) }{k_{c}}\dfrac{1}{A}-\dfrac{\omega }{\pi k_{c}}\dfrac{%
1}{A^{2}}\dfrac{\partial I\left( A,\varphi \right) }{\partial \varphi }
\end{array}
\right. \text{,}  \label{equequationmouvement}
\end{equation}
with~:

\begin{equation}
I\left( A,\varphi \right) =\int_{0}^{T}V_{int}\left[ z\left( t\right) \right]
dt
\end{equation}
The system given by equs.\ref{equequationmouvement} is defined for any kind
of interaction potential and no particular hypothesis are required to
perform the calculations. This implies that, provided that $I\left(
A,\varphi \right) $ is analytical, the equations of motion can be obtained
for any kind of non-linearity. For instance, in ref.\cite{Nony00}, analogous
calculations were performed for the Duffing's oscillator. If we consider the
sphere-plane interaction involving the disperse part of the Van der Waals
potential (see equ.\ref{equpotVdWdisp}), it was shown \cite{Nony99}\ that~:

\begin{equation}
I\left( A,\varphi \right) =-\frac{\pi HR}{3\omega }\frac{1}{\sqrt{D^{2}-A^{2}%
}}
\end{equation}
Thus~:

\begin{equation}
\frac{\partial I\left( A,\varphi \right) }{\partial A}=-\frac{\pi HR}{%
3\omega }\frac{A}{\left( D^{2}-A^{2}\right) ^{3/2}}\qquad \text{and}\qquad
\frac{\partial I\left( A,\varphi \right) }{\partial \varphi }=0
\end{equation}
Using usual dimensionless notations \cite{Nony99}, the equs.\ref
{equequationmouvement} of the coupled equations of motion becomes~:

\begin{equation}
\left\{
\begin{array}{c}
\ddot{a}=\left[ \left( u+\dot{\varphi}\right) ^{2}-1\right] a-\dfrac{\dot{a}%
}{Q}+\dfrac{\cos \left( \varphi \right) }{Q}+\dfrac{a\kappa _{a}}{3\left(
d^{2}-a^{2}\right) ^{3/2}} \\
\ddot{\varphi}=-\left( \dfrac{2\dot{a}}{a}+\dfrac{1}{Q}\right) \left( u+\dot{%
\varphi}\right) -\dfrac{\sin \left( \varphi \right) }{aQ}
\end{array}
\right. \text{,}  \label{equequationmouvementadim}
\end{equation}
$d=D/A_{0}$ is the reduced distance between the location of the surface and
the equilibrium position of the OTCS normalized to the resonance amplitude $%
A_{0}=Q\mathcal{F}_{exc}/k_{c}$, $a=A/A_{0}$ is the reduced amplitude, $%
u=\omega /\omega _{0}$ is the reduced drive frequency normalized to the
resonance frequency of the free OTCS and $\kappa _{a}=HR/\left(
k_{c}A_{0}^{3}\right) $ is the dimensionless parameter that characterizes
the strength of the interaction. The explicit dependance of the coupling
term $\kappa _{a}$ with the oscillation amplitude through the power law $%
A_{0}^{-3}$ indicates the non-linear character of the dynamics.

\subsection{Resonance peak and amplitude variation recorded during a Tapping
experiment\label{subsectionresonancepeak}}

The equations of motion of the stationary solutions $a$ and $\varphi $ are
obtained by setting $\dot{a}=\dot{\varphi}=0$ and $\ddot{a}=\ddot{\varphi}=0$
in equ.\ref{equequationmouvementadim} and lead to two coupled equations of
the sine and cosine of the phase of the OTCS previously calculated \cite
{Nony99}~:

\begin{equation}
\left\{
\begin{array}{l}
\cos \left( \varphi \right) =Qa(1-u^{2})-\dfrac{aQ\kappa _{a}}{3\left(
d^{2}-a^{2}\right) ^{3/2}} \\
\sin \left( \varphi \right) =-ua
\end{array}
\right. \text{,}  \label{equcossin}
\end{equation}
Solving equ.\ref{equcossin} gives the relationship between the sweep
frequency and the amplitude at a given distance $d$ \cite{Aime3_99}~:

\begin{equation}
u_{\pm }\left( a\right) =\sqrt{\frac{1}{a^{2}}-\frac{1}{4Q^{2}}\left( 1\mp
\sqrt{1-4Q^{2}\left( 1-\frac{1}{a^{2}}-\frac{\kappa _{a}}{3\left(
d^{2}-a^{2}\right) ^{3/2}}\right) }\right) ^{2}}  \label{equuapprochevar}
\end{equation}
The signs plus and minus are deduced from the sign of $\cos \left( \varphi
\right) $ and correspond to values of the phase ranging from $0$ to $-90%
{{}^\circ}%
$ ($u_{-}$, $\cos \left( \varphi \right) >0$) or from $-90%
{{}^\circ}%
$ to $-180%
{{}^\circ}%
$ ($u_{+}$, $\cos \left( \varphi \right) <0$), in agreement with the sign
convention of the phase in equ.\ref{equfonctionharmonique}. From equ.\ref
{equuapprochevar} is calculated the resonance peak at any reduced distance
for a given strength of the sphere-surface interaction. The two branches
define the distortion of the resonance peak as a function of $d$. $u_{-}$
gives the evolution of the resonance peak for frequency values below the
resonance one and $u_{+}$ for frequency values above the resonance.

For the description of a Tapping experiment, the variation of the amplitude $%
a$ as a function of the distance $d$ is readily obtained by rewriting equ.%
\ref{equuapprochevar} as follow~:

\begin{equation}
d_{_{\pm }}=\sqrt{a^{2}+\left[ \frac{Q\kappa _{a}}{3\left\{ Q\left(
1-u^{2}\right) \mp \sqrt{1/a^{2}-u^{2}}\right\} }\right] ^{2/3}}
\label{equdattractif}
\end{equation}
Consequently, depending on the drive frequency and the drive amplitude
(through the $A_{0}^{-3}$ dependance into $\kappa _{a}$), bifurcations from
a stable to a bistable state may occur leading to amplitude and phase jumps.
From an experimental point of view, the conditions required for the
appearance of the bifurcations have been extensively discussed in refs.\cite
{Boisgard98, Nony99}. In particular, the use of drive frequencies lower than
the resonance frequency favor the measurement of the bifurcations
corresponding to the non-contact situations.

\begin{figure}
\includegraphics[width=10cm]{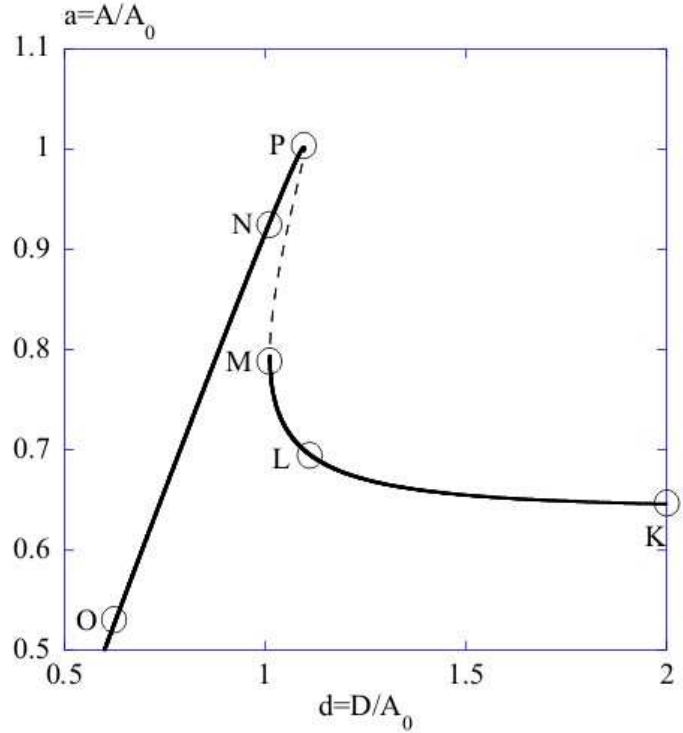}\\
  \caption{Variations of the amplitude as a function of the distance,
e.g. approach-retract curve in the tapping mode computed from equation (17). The numerical parameters are the same than in Figure 1 and the drive
frequency chosen is $u_{drive} = 0.9985$. The curve exhibits a hysteresis cycle (MNPL) due to the non-linear coupling that characterizes
bifurcations (points M and P) from a monostable to a bistable state (see text). The stable domains of the branches are shown with continuous lines
and the unstable domain with dashed line.}\label{figcartapping}
\end{figure}

In figs.\ref{figreso} and \ref{figcartapping} are given the distortion of the resonance peak and the evolution of the amplitude as a function of
$d$ for an identical set of parameters. For large values of $d$, e.g. when the surface is far from the OTCS (point K), the non-linear effects
are negligible and the peak keeps a well-defined Lorentzian shape (see equ.%
\ref{equuapprochevar} with $\kappa _{a}=0$). When the OTCS is approached
towards the surface, because the interaction is attractive, the resonance
peak starts to distort towards the low frequencies. The distortion of the
peak increases as $d$ decreases. In the vicinity of the resonance, $%
u\lesssim 1$, $a\gtrsim 1$ and for small values of $d$ the branches $u_{+}$
and $u_{-}$ become very close. To mimic a Tapping experiment, the drive
frequency is fixed to $u_{drive}$\ in fig.\ref{figcartapping}. With the
parameters used, at $d_{1}=2$ the oscillations properties are nearly
identical to those at $d$ infinite. When the peak starts to distort the
amplitude and the phase (not shown) vary (L). In the present example,
because $u_{drive}$ is chosen below the resonance, the amplitude first
increases. When the OTCS is further approached it reaches an unstable branch
(M) and jumps to the stable branch (N) such that the bistable structure of
the oscillator can be experimentally observed. At a closer distance, the
peak further distorts and the amplitude is reduced since it follows the
variations of $u_{+}$ (point O on fig.\ref{figcartapping}). Then, when the
oscillator is retracted, it follows the upper stable branch until the
resonance value corresponds to the fixed frequency $u_{drive}$ (point P on
fig.\ref{figcartapping}) and then jumps down to the lower stable branch. The
curve exhibits a hysteresis cycle (points M, N, P and L). On the figs.\ref
{figreso}\textit{\ }and \ref{figcartapping}, the branches that are supposed
unstable are shown with dashed lines.

\subsection{Resonance frequency shift}

\begin{figure}
\includegraphics[width=10cm]{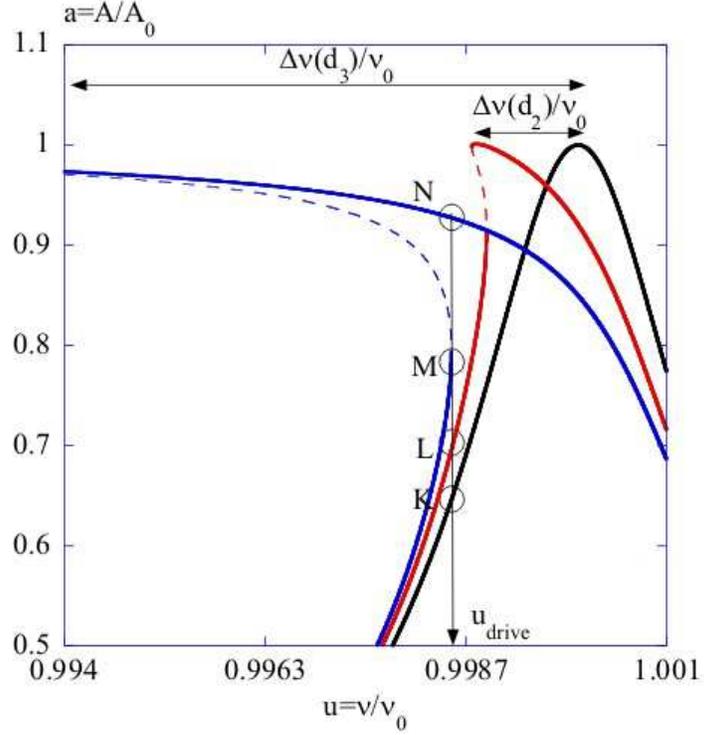}\\
  \caption{Evolution of the resonance peak computed from equation
(16) for three values of the distance, $d_1 = 2$, $d_2 = 1.11$ and $d_3 = 1.012$. The numerical parameters are $A_0 = 20$~nm, $Q = 400$ and
$\kappa_a = 8\times 10^{-4}$. For an attractive coupling, the peak is more and more distorted towards the low frequencies as $d$ is reduced, e.g.
the surface is approached and lead to the bifurcations observed on the tapping curve (Fig. 3, point M) when the drive frequency, $u_{drive} =
0.9985$, is chosen below the resonance and to the variations of the resonance frequency shift (Fig. 4). For each value of $d$, the unstable
domains of $u_-$ are shown with dashed lines.}\label{figreso}
\end{figure}

\begin{figure}
\includegraphics[width=10cm]{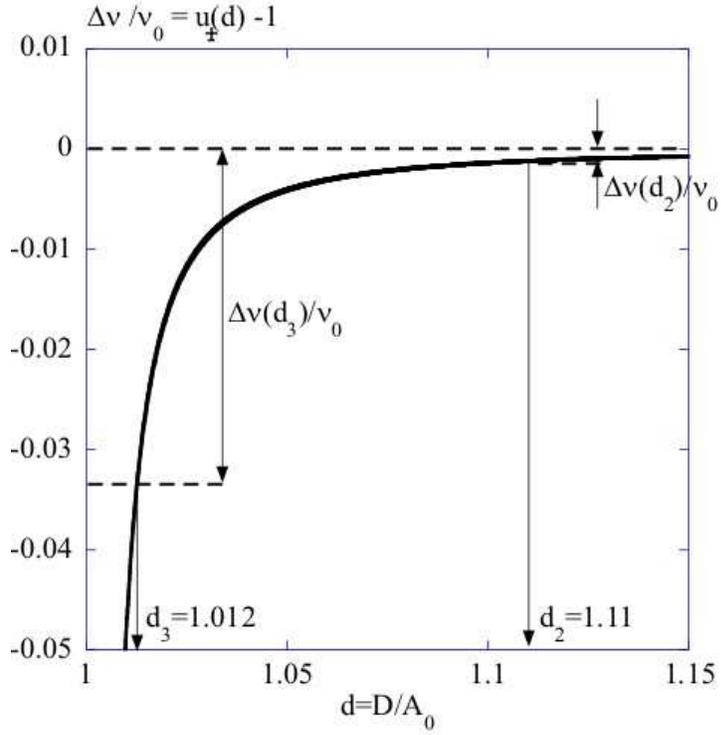}\\
  \caption{Variations of the frequency shift of the resonance peak
of the oscillator as a function of the distance, e.g. approach-retract curve in NC-AFM mode computed from equation (18). The numerical parameters
are the same than in Figure 1. It's predicted that the curve is stable with $d$.}\label{figcarncafm}
\end{figure}

Using equ.\ref{equuapprochevar},\ the resonance frequency shift as a
function of the distance $d$ is obtained by setting $a=1$. This former
condition ensures the required condition for the NC-AFM mode. Thus, the
normalized frequency shift, $\left( \nu -\nu _{0}\right) /\nu _{0}$,\ is
given by $u-1$ \cite{Aime3_99}~:

\begin{equation}
u_{_{\pm }}\left( d\right) -1=\sqrt{1-\frac{1}{4Q^{2}}\left( 1\mp \sqrt{1+%
\frac{4}{3}\frac{Q^{2}\kappa _{a}}{\left( d^{2}-1\right) ^{3/2}}}\right) ^{2}%
}-1  \label{equshiftattractif}
\end{equation}
In fig.\ref{figcarncafm}\textit{\ }is given the frequency shift as a
function of the distance $d$ for the same set of parameters than the one of
figs.\ref{figreso}\textit{\ }and \ref{figcartapping}. Following the previous
discussion about the stability of the different parts of the resonance peak
during the distortion, since the measure is performed as a function of $d$
with $a=1$, no bistable behavior can be observed. The OTCS follows the same
branch $u_{-}$ or $u_{+}$ whose location is always stable whatever the peak
distortion.

Nevertheless, note that it should exist only one branch of variation for the
frequency shift which is defined from the condition $a=1$. But two branches
are obtained as a consequence of the two solutions $d_{_{\pm }}$. When the
peak is distorted, the branches $u_{-}$ and $u_{+}$ become very close as $Q$
becomes large (see for instance, fig.\ref{figreso} with $d_{3}=1.012$).
Therefore even with an oscillation amplitude kept constant, question rises
about the ability of the OTCS to remain on the same branch. Qualitatively,
one may expect that around $a\cong 1$,\ the branch $u_{-}$ is unstable and $%
u_{+}$ is stable (see fig.\ref{figreso}). If this is true, any small
fluctuation of the oscillation amplitude might produce a jump from one
branch to the other one as discussed in ref.\cite{Aime3_99}. Since the
branch $u_{-}$ seems to be unstable, a jump to this branch should lead to an
abrupt decrease of the amplitude, which in turn might produce an apparent
abrupt decrease of the quality factor. Because such a jump should show
accidents, both on the resonance frequency shift curve and on the damping
signal, accidents which are, in most cases, not observed, it becomes useful
to determine more accurately the stability of the two branches.

\bigskip

The main aim of the variational method is to define a theoretical frame
describing at the same time the evolution of all the variables~: $a$, $%
\varphi $, $u$ and $d$ of the OTCS dynamics. Thus, it becomes useless to
discuss about the stability of each kind of branch for the two dynamic modes
since the stability of a given couple $\left( a,\varphi \right) $, or $%
\left( a,u\right) $ as a function of $d$ can be deduced for any other
couple. The discussion will be made on the branches of the resonance peak.

\section{Stability criterions\label{sectionstability}}

The stability of the branches $u_{\pm }$ of the resonance peak (see equ.\ref
{equuapprochevar}) is obtained from equations of motion of the OTCS (see equ.%
\ref{equequationmouvementadim}). These equations are linearized around the
stationary solution which will be now identified by the index ``$s$''. At
this stage, we fall into the well-known linear theory (see for instance
refs. \cite{Manneville91} and \cite{Gutzwiller91}).\ Writing into a four
dimensions matrix the linearized system corresponding to the generalized
coordinates $\left( a,\varphi \right) $ and their associated generalized
velocities $\left( \dot{a},\dot{\varphi}\right) $, we extract the
eigenvalues and discuss the stability as a function of the sign, negative
(stable solution) or positive (unstable solution), of their real part.\ The
calculations are detailed in the appendix.

\begin{figure}
\includegraphics[width=10cm]{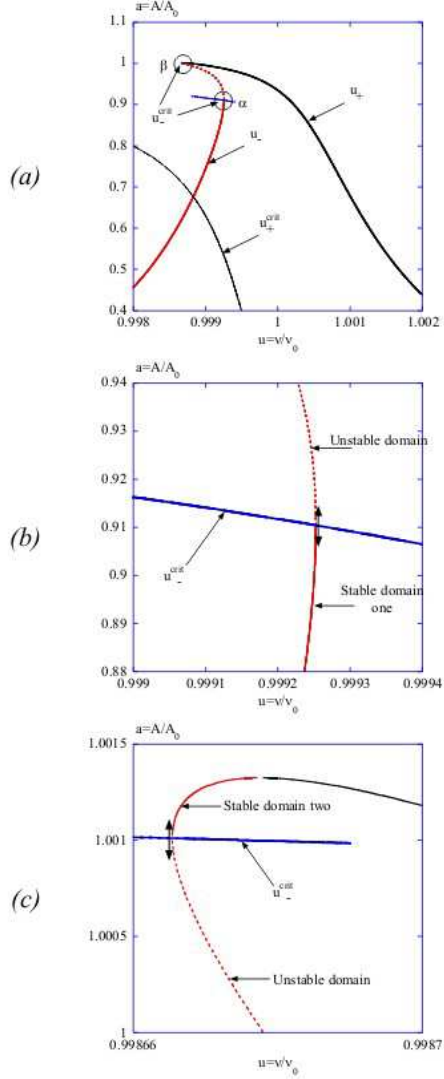}\\
  \caption{(a) Distortion of the resonance peak computed from equation (16). The numerical parameters are $d = 1.05$, $A_0 = 10$~nm, $Q = 500$ and
  $\kappa_a = 2.5\times 10^{-4}$. The critical branches $u_\pm^{crit}$ are calculated from equation (19). $u_+$ is always located above its
  critical branch $u_+^{crit}$ and in turn is always
stable. $u_-^{crit}$ crosses $u_-$ twice. This leads to define three domains to describe the stability of the branch. The domains are defined
between the spots where the derivative $da/du_-$ diverges (see text). (b) Zoom in the region $\alpha$ of $u_-$. The stability criterion foresee
that below $u_-^{crit}$, $u_-$ is stable and unstable above. This is illustrated by the dashed lines. (c) Zoom in the region $\beta$ of $u_-$. As
$da/du_-$ diverges again, it defines a new domain of $u_-$ which is predicted to be stable.}\label{figstabilitebranches}
\end{figure}

The stability criterion for each one of the two branches is given by the
following inequation (see equ.\ref{equinequstabil})~:

\begin{equation}
\left( \frac{u}{Q}\right) ^{2}-\left( u^{2}-1+\Delta \right) \frac{\cos
\left( \varphi _{s}\right) }{Qa_{s}}>0  \label{equconditionstabilite1}
\end{equation}
$\Delta $ is proportional to $\kappa _{a}$ (see appendix, equ.\ref{equDelta}%
),\ thus characterizes the non-linear attractive coupling term. Solving the
associated equality gives the critical branch $u_{\pm }^{crit}$ whose
location regardless $u_{\pm }$ defines the stability of each branch.
Unfortunately, $u_{\pm }^{crit}$ has no simple expression. Therefore the
stability condition \ref{equconditionstabilite1} can not be exploited as is.
Nevertheless a numerical routine allows to get its main features. The fig.%
\ref{figstabilitebranches}\textit{(a)}\ shows the distortion of the
resonance peak and the critical branch $u_{\pm }^{crit}$ numerically
computed\ for each branch. $u_{+}^{crit}$ never crosses $u_{+}$ and is
always located below it. On the contrary, $u_{-}^{crit}$ crosses $u_{-}$
twice and their relative position depends on the value of the amplitude
which, in turn, is going to define two domains of stability. The figs.\ref
{figstabilitebranches}\textit{(b) }and\textit{\ }\ref{figstabilitebranches}%
\textit{(c) }are zooms on the regions $\alpha $ and $\beta $ of $u_{-}$. The
intersection spots are exactly located where the curvature of $u_{-}$
changes. Therefore it's worth discussing the stability as a function of the
local curvature of the branches $u_{\pm }$ and so introducing their
derivative $da_{s}/du_{\pm }$. As shown in the appendix, the inequation \ref
{equconditionstabilite1} can be summarized as follow~:

\begin{equation}
\left\{
\begin{array}{c}
\dfrac{da_{s}}{du}>0\qquad \text{and}\qquad \cos \left( \varphi _{s}\right)
>a_{s}/\left( 2Q\right) \qquad (i) \\
\text{or} \\
\dfrac{da_{s}}{du}<0\qquad \text{and}\qquad \cos \left( \varphi _{s}\right)
<a_{s}/\left( 2Q\right) \qquad (ii)
\end{array}
\right.  \label{equconditionstabilite2}
\end{equation}
The aim of this former expression is to exhibit an explicit and particularly
simple dependence of the stability of the branches as a function of their
derivative~:

$\looparrowright $ For the $u_{+}$ branch, $da_{s}/du_{+}$ being always
negative and the associated value of the phase being always defined beyond $%
-90%
{{}^\circ}%
$ (see section \ref{subsectionresonancepeak}), thus $\cos \left( \varphi
_{s}\right) <0$, the criterion $(ii)$ implies that the $u_{+}$ branch is
always stable, whatever the value of $a_{s}$.

$\looparrowright $ Concerning $u_{-}$, the sign of the derivative changes
twice. For this branch, the phase is always defined above $-90%
{{}^\circ}%
$ which in turn means $\cos \left( \varphi _{s}\right) >0$. Therefore on the
lower part of the branch (small $a$), $da_{s}/du_{-}>0$ and the criterion $%
(i)$ indicates that the branch is locally stable. When $da_{s}/du_{-}$
becomes negative (see fig.\ref{figstabilitebranches}\textit{(b)}), because $%
\cos \left( \varphi _{s}\right) $ is still positive, the criterion $(i)$ is
no more filled. As a consequence, $u_{-}$ is locally unstable and the
instability is precisely located where the infinite tangent appears. On the
upper part of the resonance peak, the curvature of $u_{-}$ changes again and
$da_{s}/du_{-}>0$ (see fig.\ref{figstabilitebranches}\textit{(c)}), implying
that it is again a locally stable domain. Thus the branch $u_{-}$ exhibits
two stable domains and one unstable.

Note also that the resonance condition is deduced from $da_{s}/du=0$ which
implies $\cos \left( \varphi _{s}\right) =a_{s}/\left( 2Q\right) $. This
equality is the usual resonance condition of a free harmonic oscillator. If $%
a_{s}=1$, e.g. without any coupling, the resonance phase is therefore $%
\varphi _{s}=\arccos \left[ 1/\left( 2Q\right) \right] $. For the OTCS we
used, $Q\simeq 500$, and so $\varphi _{s}\cong -90%
{{}^\circ}%
$.

\section{Discussion}

The previous criterions allow to conclude to the stability of the OTCS for
each dynamic mode. It was shown that, for an attractive coupling, the branch
$u_{+}$ was always stable and that the instability was controlled by $u_{-}$%
. For frequencies lower than the resonance (branch $u_{-}$), when the
tangent $da_{s}/du_{-}$ is positive, the branch is stable. Thus, there is a
small domain close to the resonance value for which the $u_{-}$ branch
remains stable.

\bigskip

For the Tapping, this result implies that the OTCS is locked on a stable
branch until $da_{s}/du_{-}\rightarrow \pm \infty $ (see figs.\ref{figreso}
and \ref{figcartapping}, point M) which makes the amplitude jumping up to
the upper stable branch $u_{+}$ (point N on fig.\ref{figcartapping}) during
the approach and jumping down to the lower stable one $u_{-}$ (point P)
during the retract.

\bigskip

For the NC-AFM, the result of the present work shows that $u_{+}$ is always
stable but that also a small domain of $u_{-}$ around the resonance value
remains stable. If the resonance value would have been located at the point
where $da_{s}/du_{-}$ is infinite, an infinitely small fluctuation would
have been able to generate a catastrophic behavior like large variations of
the oscillation amplitude and lead to an abrupt increase of the damping
signal as previously discussed and suggested in ref.\cite{Aime3_99}.

\begin{figure}
\includegraphics[width=10cm]{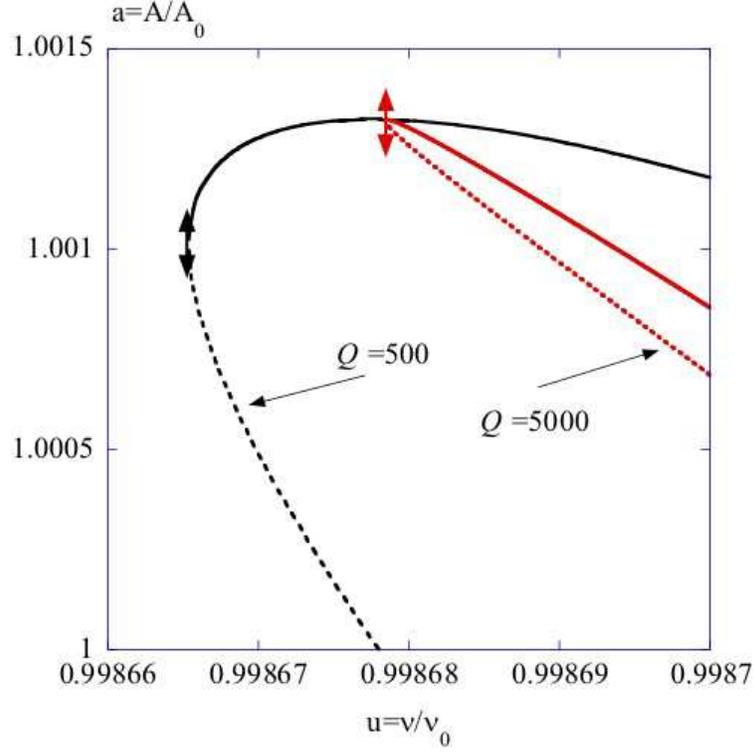}\\
  \caption{Zoom on the same scale than Figure 5c of the distortion of the resonance peak for two values of the quality factor of the OTCS, $Q = 500$ and $Q =
5000$. The numerical parameters are the same than in Figures 5. The unstable domain of $u_-$ is still shown with dashed lines but the size of the
second stable domain is drastically reduced for the larger value of $Q$ so that it nearly no appears on this scale. The associated phase
variations regardless $-90^\circ$ (not shown) are of about $1.5^\circ$ and $0.15^\circ$ for $Q = 500$ and $5000$, respectively (see
text).}\label{figqeffect}
\end{figure}

Nevertheless, the size of the $u_{-}$ stable domain is $Q$ dependant. The
more the $Q$ factor is large, the more the size of the domain is reduced.
The fig.\ref{figqeffect}\ illustrates the reduction of the size of the
domain for $Q=500$ and $Q=5000$. With $Q=5000$, the size of the domain is so
weak that it nearly no appears on the scale of the figure. As a consequence,
even if the previous discussion may help to understand why abrupt increases
of the damping signal do not systematically occur, question remains unclear
for the very high $Q$ factors that can be obtained in ultra-high vacuum $%
\left( Q\gtrsim 10000\right) $. At this step, it's worth giving orders of
magnitude. During a NC-AFM experiment, an electronic feedback loop keeps
constant the amplitude of the OTCS and locks its phase at $-90%
{{}^\circ}%
$. Therefore question rises about the size of the stable domain in phase
around $-90%
{{}^\circ}%
$. If any fluctuation around the locked value goes beyond the stable domain,
the OTCS behavior becomes unstable. For $Q=500$, the size of the stable
domain is of about $1.5%
{{}^\circ}%
$ whereas it's reduced to $0.15%
{{}^\circ}%
$ for $Q=5000$ (data not shown). Thus, if the electronic loop is able to
control the phase locking with a better accuracy than $0.15%
{{}^\circ}%
$, the OTCS will be locked within a stable domain and in turn won't give rise
to instabilities. In addition, since the ability of the electronic loop to
control the oscillating behavior depends on the value of the quality factor
\cite{Couturier01}, the reduction of the domain of stability might not be a
key parameter.

Practically, during our experiments, even with quality factors larger than $%
10000$, drastic variations of the oscillation amplitude are never observed.
Thus, the main aim of the present work is to show that, if the oscillator is
properly locked at the $-90%
{{}^\circ}%
$ value throughout an experiment, this value corresponds to a stable domain.

\section{Conclusion}

This paper was a theoretical investigation of the stability of the
non-linear behavior of an oscillating tip-cantilever system close to a
surface. A variational principle has allowed to get the temporal dependance
of equations of motion of the oscillator as a function of the non-linear
attractive coupling. The interaction potential chosen is a disperse Van der
Waals one, calculated between a sphere and a plane. The stationary state is
obtained and can be interpreted either in the Tapping mode or in the NC-AFM
mode. The stability of the stationary state is analyzed in terms of
distortion of the resonance peak as a function of the coupling. It is found
that stability criterions can be expressed from a simple inequality
involving the sign of the derivative of the curve. The branch associated to
the frequencies larger than the resonance is always stable whereas the
branch associated to the frequencies smaller than the resonance exhibits two
stable domains. The instability appears when the branch exhibits an infinite
tangent. This feature allows to re-interpret the instabilities appearing in
Tapping mode and may help to understand why the NC-AFM mode is stable most
of time.

\section*{Appendix: Computation of the stability of the branches\label{Appendix}}
Let's note $a=a_{s}\left( 1+\xi \right) $ and $\varphi =\varphi _{s}+p$ with $\xi ,p\ll 1$. The index $s$ is attributed to the stationary
solution. Keeping the terms of first order in equ.\ref{equequationmouvementadim},
equations of motion of the variations $\xi $ and $p$ regardless $a_{s}$ and $%
\varphi _{s}$ respectively, may be written as~:

\begin{equation}
\left\{
\begin{array}{c}
\ddot{\xi}=\left( u^{2}-1+\Delta \right) \xi +\dfrac{u}{Q}p-\dfrac{1}{Q}\dot{%
\xi}+2u\dot{p} \\
\ddot{p}=-\dfrac{u}{Q}\xi -\dfrac{\cos \left( \varphi _{s}\right) }{Qa_{s}}%
p-2u\dot{\xi}-\dfrac{1}{Q}\dot{p}
\end{array}
\right. \text{,}
\end{equation}
with~:

\begin{equation}
\Delta =\frac{\kappa _{a}}{3\left( d^{2}-a_{s}^{2}\right) ^{3/2}}\left( 1+%
\frac{3a_{s}}{d^{2}-a_{s}^{2}}\right)  \label{equDelta}
\end{equation}
The system is solved by setting $\Xi =\dot{\xi}$, and $\Psi =\dot{p}$
leading to a linear system of the fourth order~:

\begin{equation}
\left(
\begin{array}{c}
\dot{\xi} \\
\dot{p} \\
\dot{\Xi} \\
\dot{\Psi}
\end{array}
\right) =\Bbb{M}\left(
\begin{array}{c}
\xi \\
p \\
\Xi \\
\Psi
\end{array}
\right) \text{,}
\end{equation}
with~:

\begin{equation}
\Bbb{M=}\left(
\begin{array}{cccc}
0 & 0 & 1 & 0 \\
0 & 0 & 0 & 1 \\
u^{2}-1+\Delta & u/Q & -1/Q & 2u \\
-u/Q & -\cos \left( \varphi _{s}\right) /\left( Qa_{s}\right) & -2u & -1/Q
\end{array}
\right)
\end{equation}
The eigenvalues of the matrix are obtained by solving the characteristic
polynom given by $P=det\left( \Bbb{M}-\lambda \Bbb{I}\right) $. $P$ can then
be written as~:

\begin{equation}
P=\left( \lambda ^{2}+\lambda /Q+M\right) \left( \lambda ^{2}+\lambda
/Q+N\right)
\end{equation}
By identification~:

\begin{equation}
\left\{
\begin{array}{c}
M+N=3u^{2}+1-\Delta +\dfrac{\cos \left( \varphi _{s}\right) }{Qa_{s}} \\
MN=\left( u/Q\right) ^{2}-\left( u^{2}-1+\Delta \right) \dfrac{\cos \left(
\varphi _{s}\right) }{Qa_{s}}
\end{array}
\right.  \label{equconditionssurMetN}
\end{equation}
The characteristic equation $P=0$ is then equivalent to the following
system~:

\begin{equation}
\left\{
\begin{array}{c}
\lambda ^{2}+\lambda /Q+M=0 \\
\lambda ^{2}+\lambda /Q+N=0
\end{array}
\right. \Leftrightarrow \left\{
\begin{array}{c}
\lambda _{1,2}=\left( -1/Q\pm \sqrt{\left( 1/Q\right) ^{2}-4M}\right) /2 \\
\lambda _{3,4}=\left( -1/Q\pm \sqrt{\left( 1/Q\right) ^{2}-4N}\right) /2
\end{array}
\right.
\end{equation}
The stable solutions are the ones given by $\Re \left( \lambda _{i}\right)
<0 $ \cite{Manneville91}, thus~:

\begin{equation}
\left\{
\begin{array}{c}
1/Q>\sqrt{\left( 1/Q\right) ^{2}-4M} \\
\text{and} \\
1/Q>\sqrt{\left( 1/Q\right) ^{2}-4N}
\end{array}
\right. \Leftrightarrow M>0\hspace{0.5cm}\text{and}\hspace{0.5cm}N>0
\label{equMetNpositifs}
\end{equation}
According to relationships \ref{equconditionssurMetN}, two conditions are
necessary to fill equ.\ref{equMetNpositifs} and in turn to ensure\ the
stability of the solutions~: $MN>0$ and $M+N>0$.

$\smallskip \looparrowright $Let's first consider $MN>0$ which is the main
of the both (see below)~:

\begin{equation}
MN=\left( \frac{u}{Q}\right) ^{2}-\left( u^{2}-1+\Delta \right) \frac{\cos
\left( \varphi _{s}\right) }{Qa_{s}}>0  \label{equinequstabil}
\end{equation}
The equation $MN=0$ can be numerically solved using a Maple routine.
Nevertheless a tractable stability criterion requires to write in a
different way the expression \ref{equinequstabil}. Using the relationship $%
\cos \left( \varphi _{s}\right) =\pm \sqrt{1-\left( ua_{s}\right) ^{2}}$
(see equ.\ref{equcossin}), the two coupled equations of the sine and cosine
of the phase of the stationary state imply~:

\begin{equation}
G\left( a_{s},u\right) =Qa_{s}\left( 1-u^{2}\right) -g\left( a_{s}\right)
\mp \sqrt{1-\left( ua_{s}\right) ^{2}}=0\text{,}
\end{equation}
with~:

\begin{equation}
g\left( a_{s}\right) =\frac{a_{s}Q\kappa _{a}}{3\left(
d^{2}-a_{s}^{2}\right) ^{3/2}}
\end{equation}
Therefore~:

\begin{equation}
dG\left( a_{s},u\right) =\partial _{a_{s}}G\left( a_{s},u\right)
da_{s}+\partial _{u}G\left( a_{s},u\right) du=0\text{,}
\end{equation}
and so~:

\begin{equation}
\frac{da_{s}}{du}=-\frac{\partial _{u}G\left( a_{s},u\right) }{\partial
_{a_{s}}G\left( a_{s},u\right) }
\end{equation}
The calculations lead to~:

\begin{equation}
\frac{da_{s}}{du}=\dfrac{u}{Q}\times \frac{2\cos \left( \varphi _{s}\right)
-a_{s}/Q}{\left( \dfrac{u}{Q}\right) ^{2}-\left( u^{2}-1+\Delta \right)
\dfrac{\cos \left( \varphi _{s}\right) }{Qa_{s}}}
\end{equation}
The denominator is exactly the product $MN$, thus~:

\begin{equation}
MN>0\Leftrightarrow \dfrac{u}{Q}\times \frac{2\cos \left( \varphi
_{s}\right) -a_{s}/Q}{\dfrac{da_{s}}{du}}>0
\end{equation}
$u/Q$ being always positive, the stability condition $MN>0$ is reduced to~:

\begin{equation}
\frac{2\cos \left( \varphi _{s}\right) -a_{s}/Q}{\dfrac{da_{s}}{du}}%
>0\Leftrightarrow \left\{
\begin{array}{c}
\dfrac{da_{s}}{du}>0\qquad \text{and}\qquad \cos \left( \varphi _{s}\right)
>a_{s}/\left( 2Q\right) \\
\text{or} \\
\dfrac{da_{s}}{du}<0\qquad \text{and}\qquad \cos \left( \varphi _{s}\right)
<a_{s}/\left( 2Q\right)
\end{array}
\right.  \label{equcriterestabilite}
\end{equation}

\smallskip

$\looparrowright $Let's now consider the condition $M+N>0$~:

\begin{equation}
M+N=3u^{2}+1-\Delta +\dfrac{\cos \left( \varphi _{s}\right) }{Qa_{s}}>0
\end{equation}
Considering that $a_{s}$ varies within the range $\left[ 0..1+\varepsilon %
\right] $, with $1\gg \varepsilon >0$, it's straightforward to show that the
inequality is always filled so that the stability criterions are only given
by the condition $MN>0$ and equ.\ref{equcriterestabilite}.

\section*{References}

\bibliographystyle{unsrt}

\section*{Figures}

\end{document}